\RequirePackage{amsmath}                      

\documentclass{vgtc}                          

\graphicspath{{figures/}{pictures/}{images/}{./}} 

\usepackage{times}                     

\usepackage{tabu}                      
\usepackage{booktabs}                  
\usepackage{lipsum}                    
\usepackage{mwe}                       

\usepackage{mathptmx}                  

\usepackage{xcolor}
\usepackage{environ}

\newif\ifshowchanges
\showchangesfalse

\ifshowchanges
  \newcommand{\added}[1]{{\color{blue}#1}}
  \newcommand{\deleted}[1]{{\color{red}#1}}
  \NewEnviron{addedblock}{\begingroup\color{blue}\BODY\endgroup}
  \NewEnviron{deletedblock}{\begingroup\color{red}\BODY\endgroup}
\else
  \newcommand{\added}[1]{#1}
  \newcommand{\deleted}[1]{}
  \newcommand{\editnote}[1]{}
  \NewEnviron{addedblock}{\BODY}
  \NewEnviron{deletedblock}{}
\fi

\usepackage{booktabs}
\usepackage{tabularx}
\usepackage{array}
\usepackage{xcolor}

\newcolumntype{Y}{>{\raggedright\arraybackslash}X}

\definecolor{dpbetter}{RGB}{0,80,180}
\definecolor{dpworse}{RGB}{180,30,30}

\newcolumntype{L}[1]{>{\raggedright\arraybackslash}p{#1}}
\onlineid{1280}

\vgtccategory{Theory or Model}

\vgtcinsertpkg

\title{Rethinking Utility in Differentially Private Visualization: A Channel-Centric Perspective}

\author{Jing Yang\thanks{e-mails: \{jing.yang, spacca, zyang19, liyue.fan\}@charlotte.edu} %
\and Sarayu Pacca %
\and Zhaocong Yang %
\and Liyue Fan}
\affiliation{\scriptsize The University of North Carolina at Charlotte}

\abstract{
Differential privacy (DP) is increasingly adopted to protect sensitive data in visualization, but understanding how privacy-induced noise affects visual representations remains fragmented. Prior work has examined the impact of DP noise on specific \deleted{visualization}\added{chart} types, yet these findings are typically tied to particular \deleted{encodings}\added{chart types} and evaluation settings, making it difficult to generalize across contexts. In parallel, visualization research has long established that visual channels, \deleted{such as position, length, angle, and color,}\added{the visual properties of marks that encode data attributes}, differ in perceptual effectiveness under noise-free conditions. However, these principles have not been systematically revisited under DP, where noise is deliberately introduced into the data to protect privacy.

In this paper, we present a channel-centric perspective to reframe \deleted{prior empirical and theoretical results}\added{prior empirical, computational, and theoretical work}. \deleted{The synthesis surfaces patterns that channel-related choices, both within an encoding and across encodings, carry systematic utility consequences under DP. Encodings that emphasize relative or aggregate structure (proportions, spatial groupings, density patterns) tend to be more robust than those requiring precise value reading.}\added{Our synthesis draws on prior findings to motivate a channel-read-mechanism hypothesis: the utility of DP visualizations may depend on the task-relevant reads made salient by visual channel configurations and DP mechanism choice jointly.} \deleted{Moreover, this perspective shows that channel choice is an active determinant of the utility of privacy-preserving visualization, shaping how disclosure is framed, how DP algorithms are designed and selected, how utility is evaluated, and how analysts interpret noisy output.}\added{Moreover, this perspective highlights the role of channel choice throughout the privacy-preserving visualization pipeline, from disclosure framing and DP algorithm design to utility evaluation and interpretation of noisy outputs.} The synthesis also \deleted{suggests a candidate explanation for the robustness pattern, }\deleted{surfaces}\added{identifies} gaps in the existing evaluation literature and highlights research opportunities for channel-aware DP visualization.%
}


\usepackage{amsmath}

\begin{document}


\firstsection{Introduction}

\maketitle

\label{sec:introduction}

Differential privacy (DP) has become a \deleted{state-of-the-art}\added{widely adopted} paradigm for protecting sensitive data in statistical analysis and visualization~\cite{Cummings2024Advancing}. It ensures that the output of an analysis remains statistically indistinguishable when any single individual's data is added or removed~\cite{DR14}. More formally, given any neighboring databases $\mathcal{D}$ and $\mathcal{D}'$ that differ by at most one record, a randomized mechanism $\mathcal{A}$ satisfies \textbf{($\epsilon, \delta$)-DP} if for any subset ${S}  \subset range(\mathcal{A})$,
\begin{equation}\rm{Pr}[\mathcal{A}(D) \in {S}] \leq e^{\epsilon} \cdot \rm{Pr}[\mathcal{A}(D') \in {S}] + \delta.
\end{equation}
The parameters $\epsilon\ge0$ and $\delta\ge0$  specify the degree of privacy provided by the mechanism.  Smaller $\epsilon$ and $\delta$ values indicate stronger privacy protection, and vice versa.  For aggregate queries, DP guarantees can be achieved by adding random noise into query results, such as through the Laplace mechanism and the Gaussian mechanism~\cite{DR14}. The noise magnitude depends on the query's sensitivity and privacy parameters (smaller $\epsilon$ and $\delta$ indicate stronger privacy and may yield larger noise). Most existing studies on DP visualization~\cite{GHK+23,PCA+24,ZSM21} adopt the $\epsilon$-DP model (i.e., setting $\delta=0$). They also adopt algorithmic approaches such as data aggregation and transformation to preserve utility while providing DP guarantees.  

\begin{deletedblock}
Visualization research has shown that visual channels (e.g., position, length, angle, and color/intensity), the perceptual dimensions through which data attributes are encoded into marks \cite{Mun14}, differ in perceptual effectiveness. Cleveland and McGill~\cite{CM84} ranked visual encodings by perceptual accuracy under noise-free conditions, and Saket et al.~\cite{Sak+18} empirically confirmed that bar charts and scatterplots are the most accurate and fastest chart types for most tasks across ten analytical task types.

Zhang et al.~\cite{ZSM21} revisited these rankings under DP, conducting a task-based evaluation of visualizations with privacy noise. Using Saket et al.'s experimental design with four chart types and eight tasks from the Amar et al. \cite{Amar2005} taxonomy, they found that \emph{summary tasks} (e.g., Find Clusters) are more noise-tolerant than \emph{value tasks} (e.g., Retrieve Value). They also observed that pie charts offer similar or better accuracy than bar charts under noise, a finding that diverges from classical rankings. However, their analysis remained at the task level: they characterized which task-visualization combinations tolerate noise, but did not examine \emph{why} certain encodings gain or lose advantage. Subsequent work on privacy-preserving visualization has expanded the design space~\cite{GHK+23,MKK25,PCA+24,RD25}, motivating a need to connect these findings and understand their broader implications.

In this paper, we synthesize Zhang et al.'s results~\cite{ZSM21} together with subsequent empirical and theoretical work~\cite{CSM+26,GHK+23,MKK25,NKW+24,PCA+24,PSD+25,RD25,RMC+19,TS25,TSP+24,TBG+22,ZWW+23} through a channel-centric lens. This synthesis highlights three patterns less visible when studies are considered in isolation. First, within-encoding design choices (bin resolution, component allocation, rendering geometry) substantially shape utility at fixed $\epsilon$. Second, channel-task effectiveness rankings shift, and in some cases flip, under DP, a pattern for which we offer a candidate explanation, developed in Section~\ref{sec:discussion}: encodings that support relative or aggregate judgments (proportions, spatial groupings, density patterns) preserve recoverable signal under DP noise, while encodings that require precise value extraction degrade more sharply. Third, channel choice reshapes practice across the DP-visualization pipeline, from disclosure framing through interpretation.

This paper makes the following contributions: (1)~a channel-centric synthesis of prior work that yields three concrete results: within-encoding design choices substantially shape utility at fixed $\epsilon$, channel-task effectiveness rankings shift and in some cases flip under DP, and channel-aware reasoning extends across the DP-visualization pipeline; (2)~a candidate explanation for the channel-task effectiveness ranking shifts; and (3)~evaluation gaps and research opportunities that motivate channel-aware evaluation and design in privacy-preserving visualization.
\end{deletedblock}

\begin{addedblock}
Building on Saket et al.'s~\cite{Sak+18} task-based evaluation of chart
effectiveness under noise-free conditions, Zhang et al.~\cite{ZSM21}
revisited chart-task combinations under DP (via Laplace mechanism) and found that
utility degradation varied substantially across chart-task pairs,
suggesting that the privacy mechanism does not affect visualizations uniformly.
Subsequent work in DP visualization has expanded the design space
by showing that choices such as rendering (e.g., bin size and geometry),
DP mechanism, and disclosure tactics also influence visualization utility
\cite{GHK+23,MKK25,PCA+24,RD25}. Together, these findings
motivate a need to connect evidence across studies and examine it
through a common lens.

Visualization research has long established that visual channels (e.g., position, length, angle, and color/intensity), the perceptual dimensions used to encode data attributes into visual marks~\cite{Mun14}, differ in perceptual effectiveness~\cite{CM84}. Because chart types are constructed from combinations of visual channels, a channel-centric perspective provides a natural lens for connecting findings across chart types and reasoning about their visual utility under DP. However, this perspective has not yet been used to synthesize findings across DP visualization studies.

In this paper, we synthesize prior empirical\added{, computational,} and theoretical work on DP visualization~\cite{CSM+26,GHK+23,MKK25,NKW+24,PCA+24,PSD+25,RD25,RMC+19,TS25,TSP+24,TBG+22,ZSM21,ZWW+23} through a channel-centric lens. We identified relevant work using Zhang et al.~\cite{ZSM21}, Panavas et al.'s study~\cite{PCA+24}, and the survey of Bhattacharjee et al.~\cite{BCD20} as starting points for backward and forward citation snowballing, supplemented by independent keyword searches (e.g., \emph{differential privacy}, \emph{visual utility}). We prioritized studies examining visualization utility under DP.

When viewed collectively through this channel-centric lens, prior studies motivate the channel-read-mechanism hypothesis for DP visualization utility: utility may partially depend on the
interaction between the task-relevant reads made salient by visual
channel configurations and the information preserved by the adopted
DP mechanisms. We further show that
channel choice matters across the DP-visualization pipeline, from
disclosure framing and DP algorithm design to utility evaluation and
interpretation.


This paper makes three contributions: (1) a channel-centric synthesis of prior empirical, computational, and theoretical work on DP visualization; (2) the channel-read-mechanism hypothesis, which posits that DP visualization utility depends in part on the interaction among visual channel configurations, the task-relevant reads they make salient, and DP mechanism choice; and (3) gaps and research opportunities for channel-aware evaluation and design in DP visualization.
\end{addedblock}

\section{Evaluation Context: A Utility Hierarchy}
\label{sec:background}

Evaluating the utility of privacy-preserving visualizations is itself a multi-faceted problem. Bhattacharjee et al.~\cite{BCD20} identify a fundamental lack of consensus on how to quantify this utility. We characterize three complementary evaluation lenses used in prior work, organized along a hierarchy from low-cost automated methods to high-cost human evaluation.

\emph{Visual utility} measures how closely a private visualization preserves features relevant for downstream tasks, typically through automated metrics: image-similarity measures or task-specific distribution scores. 
\deleted{Panavas et al.~\cite{PCA+24} benchmarked five automated metrics against expert-perceived visual quality for DP binned scatterplots (a heatmap-style encoding); MS-SSIM (multi-scale structural similarity) emerged as the strongest predictor, while standard DP metrics including APQE (average per-query error) and Earth Mover's Distance (EMD), which respects spatial proximity between bins where Manhattan distance and Kullback-Leibler divergence do not, correlated negatively for this encoding. Ryan et al.~\cite{RMC+19} validated Pixel Approximate Entropy (PAE) for shape identification accuracy in noisy line charts, though their experiments did not evaluate noise introduced by DP methods specifically.}
\added{Panavas et al.~\cite{PCA+24} benchmarked automated metrics against expert-perceived visual quality for differentially private binned scatterplots; MS-SSIM (multi-scale structural similarity) emerged as the strongest predictor, while APQE (average per-query error) and Earth Mover's Distance (EMD), two standard DP utility metrics, correlated negatively with perceived visual quality. Ryan et al.~\cite{RMC+19} showed through user studies that Pixel Approximate Entropy (PAE) correlates with perceived visual complexity in line charts, though not under DP noise.} Zhang et al.~\cite{ZSM21} illustrate the latter category, introducing Peakedness, Anomaly, and Clusteredness scores tailored to three \deleted{summary}\added{analytical} tasks (Find Maximum, Find Anomalies, Cluster) to compare DP algorithms by how well their noisy output preserves the distribution features each task requires.

\emph{Task-based utility} measures whether human viewers can correctly perform analytical tasks on noisy charts. Zhang et al.~\cite{ZSM21} proposed a dichotomous assessment model that \deleted{cleanly separates}\added{distinguishes} \textit{perceptual accuracy} (did the user correctly read the noisy chart?) from \textit{perturbation accuracy} (does the noisy answer match the true answer?), enabling targeted diagnosis of where utility breaks down. Task-based evaluation is the most direct measure of utility but also the most expensive, requiring human participants for each combination of conditions. 

\emph{Perceptual utility} operates along a distinct axis: it measures whether an authorized viewer can read the chart while unauthorized viewers cannot, addressing physical observation threats rather than inference from released outputs~\cite{CSM+26,ZMW23}.

Notably, the automated metrics used at the lower end of this hierarchy already differ by chart type and associated channel: MS-SSIM for binned scatterplots~\cite{PCA+24}, PAE for line charts~\cite{RMC+19}. Whether this pattern reflects a deeper need for channel-specific evaluation is a question we return to in Section~4.

\section{\deleted{Cross-Study Findings on Differentially Private Visualization}\added{A Channel-Centric Synthesis of DP Visualization Utility}}
\label{sec:insights}

We organize our cross-study synthesis \deleted{as three preliminary findings}\added{into two themes}. Each draws on multiple studies whose connections become \deleted{visible when examined through the lens of visual channels}\added{apparent through the lens of visual channels}.

\subsection{\deleted{Finding 1: Within an Encoding, Design Choices Determine Utility, Not Just Noise Magnitude}\added{Observations of Utility Variation under DP}}

\added{Existing studies have reported  differences in visualization utility under DP across a variety of visualization settings. The evidence spans comparisons across chart types as well as within-channel design choices, including discretization, DP mechanisms, component-specific estimation, and rendering geometry.}

\deleted{
Even before comparing across encodings, DP's effects on a single encoding are not determined by noise magnitude alone. Encoding-internal Design choices such as DP mechanisms and bin resolution and aspect ratio of the rendering can substantially shape utility at a given $\epsilon$. We illustrate this in three registers: discretization parameters and and rendering geometry.
}

\added{
\textbf{Chart-task robustness.} Zhang et al.~\cite{ZSM21} evaluated how the Laplace mechanism  affects different chart-task combinations, using a user study that measured perceptual accuracy against non-private counterparts. Their results showed that perceptual accuracy degradation varied  across chart-task pairs. For example, pie charts exhibited relatively small degradation for Filter, barplots degraded the most for Find Maximum, and line charts experienced the largest degradation for Retrieve Value, Filter, and Determine Range among the evaluated chart types. Panavas et al.~\cite{PCA+24} observed a related task-level pattern for binned scatterplots, generated with 5 DP mechanisms: Cluster retained the highest visual utility under DP, whereas Distribution retained the lowest.
}

\textbf{Discretization parameters.} 
\deleted{Encoding parameters may substantially affect utility at the same $\epsilon$ level. Panavas et al.\ showed that coarser bins ($32{\times}32$) retain significantly more visual utility than finer bins ($64{\times}64$), and that color scale adjustment (gamma values) can change MS-SSIM scores dramatically: same data, same $\epsilon$, same mechanism, different visual outcome~\cite{PCA+24}. Ghazi et al.~\cite{GHK+23} provided a possible explanation: in standard DP mechanisms, each bin receives independent random noise, so finer grids accumulate more total noise relative to the signal. They propose an algorithm that accounts for spatial relationships between neighboring bins, maintaining visual quality even at high resolutions where naive approaches fail. The bin-size effect Panavas et al.\ observed is thus not a quirk of their setup but a fundamental consequence of discretization under DP: any encoding that bins data into cells will face the same trade-off between resolution and noise, making bin size an encoding design choice that directly affects utility.}
\added{
Panavas et al. showed that, holding the privacy level and DP mechanism constant, coarser bins (32$\times$32) retain significantly more visual utility than finer bins (64$\times$64)~\cite{PCA+24}. Ghazi et al.~\cite{GHK+23} suggested a possible explanation: finer grids split the same data across more bins, reducing per-bin signal while leaving DP noise magnitude largely unchanged. This lowers the signal-to-noise ratio of spatial patterns, suggesting that the bin-size effect reflects a broader discretization challenge in DP spatial visualization.
}

\added{
\textbf{DP mechanisms.} Holding the visualization fixed, prior work has shown that the choice of DP mechanism can substantially affect visualization utility. Zhang et al.~\cite{ZSM21} showed through metric-based simulations that, for histogram-based tasks, DAWA slightly outperformed Laplace for Find Anomalies, whereas Laplace better preserved task answers for Find Maximum and Cluster. For color/intensity-encoded binned scatterplots, Panavas et al.~\cite{PCA+24} used expert-rated visual utility and metric analysis to compare mechanisms, showing that different mechanisms yielded different levels of visual utility across analytical tasks. Together, these studies suggest that, even when the visual channel configuration is held fixed, different DP mechanisms can preserve different task-relevant structures, leading to different levels of visualization utility.}


\deleted{
\textbf{Component sensitivity.} Ramsay and Diaz-Rodriguez~\cite{RD25} showed that even within a single boxplot, where all components use length/position encoding, DP noise degrades different components unequally. The box (representing the median and quartiles) can be estimated accurately under DP, but the whiskers (representing the extremes of the data range) cannot, because extreme values are inherently harder to protect while achieving high utility. Their composite DPBoxplot addresses this challenge by using a different DP estimation strategy for each component, demonstrating that encoding-aware design can recover utility that a uniform approach loses.
}

\added{
\textbf{Component-specific estimation.} Ramsay and Diaz-Rodriguez~\cite{RD25} showed that, although the box and whiskers of a boxplot both use the position channel, they summarize different statistical quantities: the box represents the median and quartiles, whereas the whiskers depend on extreme quantiles. Their composite DPBoxplot therefore applies different DP estimators to those components, demonstrating that component-aware estimation can reduce utility loss relative to using a single estimator throughout the chart.}

\textbf{Rendering geometry.} Ryan et al.~\cite{RMC+19} showed that line chart geometry, independent of the underlying data, substantially shifts \deleted{how complex the rendered chart appears}\added{visual complexity, as measured by PAE}: doubling chart width reduces PAE by more than half, while doubling height increases it more than twofold\deleted{ (their Fig.~5)}. Combined with their experimental finding that shape identification accuracy \deleted{collapses nonlinearly with PAE}\added{drops sharply once PAE exceeds a threshold}, this implies that aspect ratio may mediate utility for private line charts:  \deleted{a wider rendering pushes the chart further from the perceptual complexity threshold beyond which trends become unrecoverable.}\added{a wider rendering may lower perceived visual complexity, helping keep the chart below the threshold at which trends become unrecoverable.} \deleted{Thaker et al.'s Overlook system~\cite{TBG+22} made the converse observation on bar/histogram length encodings: at sufficiently large data scales, DP noise can fall \emph{below} the screen's pixel resolution, rendering a noisy bar visually indistinguishable from its non-private counterpart. Where Ryan et al.\ identified a complexity ceiling above which any noise would render trends unrecoverable, Overlook identified a visibility floor below which DP noise is rendered invisible by the display itself. As with bin size for binned encodings and per-component DP design in composite encodings, these geometric parameters constitute a within-encoding design lever.}\added{Thaker et al.~\cite{TBG+22} made a complementary observation for bar/histogram length encodings: at sufficiently large data scales, DP perturbations can fall below the screen’s pixel resolution, making a noisy bar visually indistinguishable from its non-private counterpart. Together with Ryan et al.’s PAE result, this shows that rendering geometry can act as a within-chart design lever.}

\begin{table*}[t]
\centering
\caption{\added{Observations motivating the channel--read--mechanism hypothesis. PP = post-processing; GT = Geometric Truncated; DAWA/MWEM/AGrid/AHP = DP mechanisms used in~\cite{PCA+24,ZSM21}.}}
\label{tab}
\small
\renewcommand{\arraystretch}{1.08}
\setlength{\tabcolsep}{3pt}
\begin{tabular}{L{0.10\linewidth} L{0.11\linewidth} L{0.14\linewidth} L{0.14\linewidth} L{0.14\linewidth} L{0.12\linewidth} L{0.18\linewidth}}
\toprule
Source & Chart & Task & Task-relevant channel(s) & Task-relevant read & DP mechanism & Key observation \\
\midrule

Zhang et al.~\cite{ZSM21} 
& Lineplot 
& Determine Range 
& Connected position 
& Extremes (min/max) 
& Laplace + PP 
& Largest degradation among evaluated charts. \\

Zhang et al.~\cite{ZSM21} 
& Barplot 
& Find Maximum 
& Length 
& Absolute magnitude 
& Laplace + PP 
& Greater degradation than other evaluated chart types for this task. \\

Zhang et al.~\cite{ZSM21} 
& Piechart 
& Filter 
& Arc length, central angle, and
segment area jointly 
& Part--whole 
& Laplace + PP 
& Relatively small degradation. \\

Zhang et al.~\cite{ZSM21} 
& Histogram 
& Max; Anomaly; Cluster 
& Length 
& Peak; Anomaly; Cluster 
& Laplace; DAWA; MWEM 
& DAWA better for Anomaly at a lower privacy budget; Laplace better for Max and Cluster. \\

Panavas et al.~\cite{PCA+24}
& Binned scatterplot
& Cluster; Correlation; Distribution
& Color/intensity
& Dense regions; Association pattern; Overall distribution
& Laplace; GT; DAWA; AGrid; AHP 
& Clustering retained the highest utility; Distribution the lowest.\newline Coarser binning retained higher utility than finer binning. \\

Ramsay \& Diaz-Rodriguez~\cite{RD25} 
& Boxplot 
& Center/spread; Extremes; Outliers 
& Position along a common scale
& Median/quartiles; Min/max; Outlier values
& Laplace; unbounded; JointExp 
& Different estimators improved different components. \\

\bottomrule
\end{tabular}
\end{table*}


\begin{addedblock}

The studies reviewed above collectively suggest that tasks, visual channels, task-relevant quantities, and DP mechanisms jointly shape visualization utility. We use \emph{read} to refer to the task-relevant quantity a viewer seeks to recover from a visualization. Table \ref{tab} summarizes these observations together with our interpretation of task-relevant channel configurations and task-relevant reads made salient by them. Note that charts are often multi-channel encodings and we list the task-relevant channel(s) following canonical channel vocabulary~\cite{CM84,Mun14}. Piecharts are represented by arc length, central angle, and segment area jointly because prior studies suggest that all three cues can affect proportional reading~\cite{Eel26,SK16}. Collectively, these observations motivate the hypothesis that visualization utility depends in part on the interaction among the visual channel configuration, the task-relevant read, and the adopted DP mechanism. We caution that these observations are derived primarily from a small number of studies. Section 4.1 develops this hypothesis.


\end{addedblock}

\begin{deletedblock}

\deleted{
\subsection{Finding 2: Channel-Task Effectiveness Rankings Shift Under DP}
}

Zhang et al.'s~\cite{ZSM21} task-based DP study tested four chart types from Saket et al.'s~\cite{Sak+18} clean-data baseline. Comparing the two studies reveals that several established rankings shift under DP, in some cases reversing outright, in others producing significant differentiation where clean-data settings showed none. We supplement Zhang's findings with channel-specific evidence from Ryan et al.~\cite{RMC+19} (slope) and Panavas et al.~\cite{PCA+24} (color/intensity), organized by channel below.

\textbf{Slope encoding collapses for distribution tasks.} Saket et al.\ found line charts to be \emph{significantly more accurate} than other chart types for Characterize Distribution and Correlation, tasks where trend perception matters most. Under DP, this advantage inverts completely: line charts were significantly \emph{worse} than all other chart types for Characterize Distribution~\cite{ZSM21}, going from best to worst for the same task. This is the starkest reversal, but Zhang et al.\ did not study why it occurs. Ryan et al.'s~\cite{RMC+19} Pixel Approximate Entropy (PAE), developed independently in the context of line chart perception, provides an intuition: PAE measures the unpredictability of a rendered line chart's pixel sequence, and their experiments show that shape identification accuracy collapses from near-perfect at low PAE to approximately chance at high PAE, with a nonlinear threshold effect. Since DP noise may directly increase a line chart's PAE by introducing discontinuities into the continuous contour that slope encoding depends on, the reversal has a quantitative explanation: noise pushes line charts past a perceptual complexity threshold beyond which trends become unrecoverable.

\textbf{Length encoding becomes noise-sensitive for precise-value tasks.} Saket et al.\ found no significant accuracy differences among chart types for Determine Range, a task that requires reading individual values. Under DP, bar chart accuracy for this task degrades significantly with increasing privacy level (i.e., reducing $\epsilon$ value)~\cite{ZSM21}; the precise-value reading that length encoding supports collapses as DP noise grows.

\textbf{Angle encoding gains advantage for value retrieval.} Saket et al.\ found that bar charts, tables, and pie charts were all top performers for Retrieve Value, with no reliable differentiation among them. Under DP, pie charts achieved significantly \emph{higher} perceptual accuracy than the other chart types~\cite{ZSM21}, a differentiation that the clean-data setting did not show. Saket et al.\ themselves noted that pie charts ``may have been further helped by having colored slices with text labels,'' so the result may partly reflect label design rather than angle encoding per se. Disentangling these factors requires controlled studies that have not yet been conducted. However, the result is consistent with angle encoding's emphasis on proportional relationships, which are mathematically better preserved than absolute magnitudes under additive noise.

\textbf{Position encoding favors aggregate spatial tasks.} Saket et al.\ found no significant accuracy differences among chart types for Filter, and found pie and bar charts most accurate for Cluster. Under DP, scatterplots emerged as significantly superior for both Filter and Cluster~\cite{ZSM21}, a position-encoding advantage for aggregate spatial tasks that clean-data evaluation did not surface.

\textbf{Color/intensity preserves aggregate patterns better than fine-detail patterns.} Panavas et al.~\cite{PCA+24} evaluated 1,200 differentially private binned scatterplots, where bin counts are encoded by color/intensity. Across the three tasks they tested (Cluster, Correlation, Distribution), cluster identification retained the highest visual utility while distribution characterization retained the lowest. They attributed the asymmetry to differences in signal-to-noise: dense regions produce strong color signals that DP noise does not overshadow, while subtle density patterns require fine color discrimination that noise disrupts. 

These cases suggest a possible explanation, developed in Section~\ref{sec:discussion-1}: encodings that support relative or aggregate judgments are more noise-tolerant than those requiring absolute values.
  
\end{deletedblock}

\subsection{\deleted{Finding 3: The Channel-Centric Perspective Reshapes Practice Across the DP-Vis Pipeline}\added{Channel Considerations Across the DP-Visualization Pipeline}}

\deleted{Beyond predicting which encoding may be superior for a given task,}The channel-centric view shapes practice upstream and downstream of visualization selection. We trace its implications at four points along the DP-visualization pipeline: how disclosure is framed in design, how DP mechanisms are tuned to preserve specific structures, how utility is measured, and how analysts interpret noisy output. At each point, the channel acts not as a passive rendering decision but as an active \deleted{determinant of}\added{factor shaping} what privacy-preserving visualization can deliver. 

\textbf{Disclosure and channel as paired design choices.} Mehta et al.~\cite{MKK25} contributed a theoretical framework that treats channel as design-relevant in privacy-preserving visualization. They distinguished \emph{disclosure tactics} (data operations that change what information is available) from \emph{emphasis tactics} (rendering choices like mark types, scales, and layouts). Their Fred scenario illustrates that disclosure tactics and channels are not chosen independently: a biostatistician sharing patient clusters while hiding individuals deliberately pairs the \emph{smoothing} tactic with \emph{contour shape}, because that combination preserves cluster visibility while suppressing individual-level detail. They acknowledged that emphasis-tactic choices ``tend to be concomitant to choices about disclosure'' but left emphasis tactics out of their analytical scope. \deleted{Our channel-centric synthesis lays an empirical foundation for that choice under DP noise: Section~3.1 and 3.2 show that channel-related choices, both within and across encodings, may have  channel-specific consequences for which utility is preserved.}\added{From a channel-centric perspective, this omission points to an important design gap: under DP noise, disclosure tactics and channel choices may jointly influence which aspects of utility are preserved.}

\textbf{DP mechanism, channel, and task as joint design choices.} Ghazi et al.~\cite{GHK+23} designed a DP mechanism specifically for heatmap visualization, with EMD as the loss function the mechanism targets, chosen for its spatial-proximity properties on color/intensity-encoded heatmaps. This illustrates DP mechanism design being tailored to a target visualization rather than treated as an independent, upstream choice. Panavas et al.~\cite{PCA+24} observed a related coupling within a single channel: DAWA \cite{LHM+14}, a data-dependent mechanism that groups nearby cells into near-uniform regions, reinforces the dense-region signal that cluster identification depends on while smoothing away the subtle density variations that distribution characterization requires. \deleted{The same mechanism aligns with one task on the color/intensity channel and works against another.}\added{Their results showed that, under DAWA, cluster-identification tasks retained higher visual utility than distribution-characterization tasks.} \deleted{Whether the optimal DP mechanism depends systematically on the visual encoding, on the analytical task, or on their combination, and whether mechanisms tuned for one such pairing transfer to others, remains an empirical conjecture; no study has directly compared DP mechanisms across channels and tasks under matched conditions.}\added{Whether mechanisms tuned for one channel–task pairing transfer to others remains an open question that requires empirical evaluation.} The evidence so far suggests that selecting a DP mechanism without considering both the visual channel and the analytical task risks discarding what the visualization is meant to communicate.

\deleted{
\textbf{Evaluation metrics need empirical validation, not just theoretical channel-fit.} The choice of evaluation metric is itself a design decision that interacts with channels. Ghazi et al.~\cite{GHK+23} chose EMD for DP heatmap evaluation based on its spatial-proximity properties, a theoretical argument for channel-fitness. Panavas et al.~\cite{PCA+24} empirically benchmarked five metrics including EMD against expert visual utility ratings on heatmap-style binned scatterplots; MS-SSIM correlated most strongly with human perception, while EMD correlated negatively. Theoretical arguments for a metric's channel-fitness do not automatically translate to predictive power for perceived utility. Panavas et al.'s~\cite{PSD+25} practitioner interviews confirmed the operational consequence: practitioners reported that general DP metrics are easy to compute but not interpretable for evaluating visualization quality. Validating which channel-specific metric actually tracks human perception requires empirical work that has not been done for most visualization types.}

\added{
\textbf{Evaluation metrics need empirical validation, not just theoretical channel-fit.} The choice of evaluation metric is itself a design decision that interacts with channels and tasks. Panavas et al.~\cite{PCA+24} empirically benchmarked five automated metrics (MS-SSIM, APQE, EMD, KS-Test, and Scagnostics) against expert ratings of visual utility across multiple heatmap-style binned scatterplots analysis tasks. MS-SSIM correlated most strongly with the expert ratings, whereas APQE, despite being a standard DP statistical metric, and EMD both correlated negatively. Theoretical arguments for a metric's channel-fitness do not automatically translate to predictive power for perceived utility. This is consistent with Panavas et al.'s~\cite{PSD+25} practitioner-interview finding that metrics tailored to particular scenarios are often more useful than general metrics.}

\textbf{The interpretation bottleneck.} Nanayakkara et al.~\cite{NKW+24} found that analysts using DP outputs lost roughly nine times as much performance to misreading noisy results as to allocating the privacy budget poorly. ViP~\cite{NBH+22} illustrated the leverage of \deleted{an encoding}\added{a visualization design} choice: it discretizes the continuous Laplace noise distribution into a quantile dotplot, turning probability-mass reading into counting dots rather than estimating area under a curve. A controlled study with 16 practitioners showed analysts read CDF probabilities more accurately from the dotplot than from a numerical-error baseline. This substitution favors count-based judgment over continuous-magnitude reading. If interpretation is a primary bottleneck, \deleted{encoding}\added{visualization design} choices are where direct leverage lies.

\section{Discussion}
\label{sec:discussion}

\begin{deletedblock}
  
\subsection{A Candidate Explanation: Relative Error of Task-Relevant Reads}
\label{sec:discussion-1}

The patterns in Table~\ref{tab} raise a hypothesis: encodings that make proportional or aggregate reads salient may be more noise-tolerant than those requiring absolute-value reads, under appropriate data conditions. A candidate explanation is that how DP noise affects the relative error of the quantity a viewer is asked to recover partially accounts for the utilities. Absolute reads, such as individual values or extrema, inherit perturbation at the level of a single released value and can have large relative error when the signal is small compared with the noise. This helps explain why barplot value and range judgments can be sensitive to DP noise.

Aggregate reads may provide a source of DP noise robustness. When a task depends on a region-level signal rather than a single noisy value, the relevant signal can accumulate across multiple bins. Under independent per-bin noise, the signal of a summed region grows linearly with the number of bins $k$, while the summed noise standard deviation grows only as $\sqrt{k}$. This signal-to-noise scaling, well known in DP visualization~\cite{TBG+22}, helps explain why binned-scatterplot clustering can remain more robust than distribution characterization: dense regions produce strong aggregate color/intensity signals, whereas subtle distributional differences require finer discrimination among noisy bins.

A similar intuition may apply to proportional reads. A proportional read compares a noisy part with a noisy whole. When the relevant counts are sufficiently large and the categories are not sparse, the $\sqrt{k}$ growth of the summed noise standard deviation provides a stable reference. This offers a candidate explanation for why piechart may gain relative advantage under DP when the task encourages part-whole judgment. However, this effect is conditional: in sparse or highly imbalanced data, proportional reads may lose this advantage.


This explanation also helps account for why the same visual attribute can behave differently across tasks and encodings. Position can support an absolute value read, a lineplot trend/shape read, or a scatterplot spatial-grouping read; color/intensity can support either a single-bin value read or an aggregate density-region read. Thus, robustness under DP is not an intrinsic property of a visual attribute alone. It depends on the interaction among channel configuration, task, data regime, and the task-relevant read made salient to the viewer. We therefore present the relative error of task-relevant reads as a hypothesis for future controlled studies and formal analysis.

\end{deletedblock}

\begin{addedblock}

\subsection{The Channel-Read-Mechanism Hypothesis}
\label{sec:discussion-1}

The observations in Table~\ref{tab} motivate the \textbf{Channel-Read-Mechanism Hypothesis}: The utility of a privacy-preserving visualization may partially depend on the interaction among the visual channel configuration, the task-relevant read it makes salient, and the adopted DP
mechanism. Because different DP mechanisms may preserve different
properties of the data, visualization utility
depends in part on how well the adopted mechanism preserves the
information required by the task-relevant read. 

For lineplots, adopting the Laplace mechanism~\cite{DR14} does not guarantee preservation of ordering between adjacent data points, and may therefore alter upward or downward trends. Such disruptions can introduce contour discontinuities, increasing visual complexity as measured by PAE~\cite{RMC+19}. Prior research shows that post-processing the results of the Laplace mechanism (e.g., to enforce non-negativity) may introduce bias and different error variances among data points~\cite{fioretto2022differential}, which may explain the difficulty with retrieving magnitudes accurately in DP barplots. 
Laplace better preserved task answers for Find Maximum and Cluster in Zhang et al.'s histogram results~\cite{ZSM21}, tasks that involve identifying
dominant or dense groups. This pattern is consistent with prior work on frequent itemset mining and web mining, where Laplace-based mechanisms have been shown to preserve large counts or frequent patterns~\cite{fan2015practical,korolova2009releasing,li2012privbasis}.

The proposed hypothesis also suggests implications for within-channel design choices and a broader range of DP mechanisms. Coarsening bin resolution or performing hierarchical decomposition~\cite{GHK+23,PCA+24,TBG+22} could better preserve aggregate density and thus retain better visual utility. Ramsay and Diaz-Rodriguez's~\cite{RD25} boxplot decomposition shows that private quantile estimators may perform differently towards inner quantiles and extreme quantiles. Incorporating specific pattern constraints in the DP mechanism may improve the visualization utility in pattern retention~\cite{ZWW+23}. 

However, existing evidence is limited. Existing studies have not yet fully isolated the channel-read-mechanism interaction from confounds such as label design, visual complexity, or task familiarity. For example, Saket et al.~\cite{Sak+18} noted that pie charts may have benefited from colored slices with text labels. Thus, the hypothesis remains a conjecture awaiting controlled studies.


\end{addedblock}

\subsection{Gaps}

\deleted{
\textbf{Visual utility metrics are channel-specific and largely unvalidated.} As noted in Section~\ref{sec:background}, the automated metrics already in use differ by encoding: MS-SSIM for binned scatterplots~\cite{PCA+24}, PAE for line charts~\cite{RMC+19}. Other channels, including angle, length, and slope, lack perceptually validated automated metrics. Theoretical arguments for a metric's channel-fitness do not automatically translate to predictive power for perceived utility, as Panavas et al.~\cite{PCA+24} demonstrate empirically (Section~\ref{sec:insights}). Validating which channel-specific metric tracks human perception requires empirical work that has not been done for most visualization types.}
\added{
\textbf{Visual utility metrics are channel--task specific and largely unvalidated.}
As noted in Section~\ref{sec:background}, the automated metrics already in use differ by visualization: MS-SSIM for binned scatterplots~\cite{PCA+24}, PAE for line charts~\cite{RMC+19}. Other channel--task combinations, including those involving angle, length, and slope, lack perceptually validated automated metrics. Theoretical arguments for a metric's fitness do not automatically translate to predictive power for perceived utility, as Panavas et al.~\cite{PCA+24} demonstrate empirically (Section~\ref{sec:insights}). Validating which metrics effectively track perceived utility for specific channel--task pairs therefore requires empirical work that remains largely undone for most visualization types.}

\textbf{Task-based evidence rests on narrow conditions.} Zhang et al.'s~\cite{ZSM21} task-based user study, a primary source of observations about chart–task variation under DP, varied chart type, task, and privacy level, but held fixed several potentially consequential dimensions: within-channel design parameters such as bin resolution and chart geometry, data properties such as scale and distribution, and the choice of DP algorithm (Laplace only). Whether the documented \deleted{ranking shifts}\added{utility-degradation patterns} hold under different conditions \deleted{is unknown}\added{remains unvalidated in task-based user studies}. Length encoding has been studied at the statistical level by Hay et al.~\cite{HMM+16} across scale, distribution, domain size, and DP algorithm; whether their statistical findings translate into task or visual utility for barplots remains uncharacterized. Visualization types beyond basic charts, such as networks, treemaps, and parallel coordinates, are largely unexamined under DP, with Tao et al.'s~\cite{TSP+24} brain connectivity work a rare exception.

\textbf{Perceptual utility has been studied largely independently of DP.} BAIT~\cite{CSM+26}, a visual-illusion-based perceptual privacy approach, demonstrated uniformly high protection across chart types for shoulder surfing but found channels differ in vulnerability: angle and slope encodings are most ``glanceable.'' Whether channel effects derived under perceptual privacy align with or diverge from observations under DP has not been examined, leaving the channel-level relationship between the two privacy regimes open.

\subsection{Opportunities}

\textbf{A \deleted{channel-specific}\added{channel--task-specific} metrics suite.} The methodology Panavas et al.~\cite{PCA+24} used to validate MS-SSIM (collecting expert visual utility ratings on a corpus of DP charts, then benchmarking candidate automated metrics against those ratings) may transfer directly to other \deleted{channels}\added{channel--task pairs}. Developing such a suite would let practitioners compare \deleted{encodings}\added{visualizations} at scale during design, rather than after deployment through expensive user studies.

 \deleted{\textit{Cross-channel comparison at matched $\epsilon$.} Extending Zhang et al.'s~\cite{ZSM21} design to cover channels underrepresented in the task-based DP literature, and to vary data scale and distribution alongside encoding choice, would provide the first systematic channel evaluation under DP. Controlled user studies could further test whether the channel effectiveness pattern is explained by the relative-vs-absolute distinction (Section~\ref{sec:discussion-1}).}
\added{\textbf{Evaluation across channels and tasks at matched $\epsilon$.} Extending Zhang et al.'s~\cite{ZSM21} \added{user study} design to cover underrepresented channel--task combinations, while varying data scale, distribution, within-encoding design choices, and DP mechanisms, would provide a more systematic evaluation of visualization utility under DP. Such studies would foster a deeper understanding of how tasks, channels, task-relevant reads, and DP mechanisms interact, and could inform new designs that better preserve utility.}
 
\textbf{Cross-channel evaluation of structure-aware DP mechanisms.} Panavas et al.~\cite{PCA+24} showed that DP mechanism choice substantially affects visual utility within a fixed channel configuration, and that a single mechanism's fit can vary by task on the same channel (Section~\ref{sec:insights}). Whether such differences extend across channels and tasks, and whether mechanisms purpose-built for specific channels or tasks (e.g.,~\cite{GHK+23,RD25,TS25}) transfer to others, has not been tested under matched conditions.
 
\textbf{Multi-channel interactions.} \deleted{Real-world encodings rarely depend on a single channel. Visualizations beyond basic charts, such as treemaps, parallel coordinates, and node-link diagrams, are inherently multi-channel.}\added{Most real-world visualizations, including many basic charts, rely on multiple visual channels. More complex visualizations, such as treemaps, parallel coordinates, and node-link diagrams, often involve even richer combinations of channels.} How channels interact under DP and whether noise on one channel compounds or compensates for noise on another remain largely open. Such interactions may also complicate mechanism selection: a mechanism optimal for a chart’s primary channel and task-relevant read may degrade secondary channels or reads. It is possible that DP mechanism design for multi-channel encodings requires joint optimization across channels and reads.

\textbf{Channel behavior across privacy threat models.} Whether BAIT's~\cite{CSM+26} channel-level vulnerability findings interact with DP-induced degradation is unexplored. \deleted{A study that varies both observation threat (BAIT-style) and DP noise across channels would clarify whether the two privacy regimes reinforce or counteract each other at the encoding level.}\added{A study that varies observation threat (BAIT-style), DP noise, visual channel, and analytical task would clarify whether the two privacy regimes reinforce or counteract each other for particular channel--task combinations.}

\subsection{Limitations}

\deleted{The channel ranking shifts trace to Zhang et al.'s~\cite{ZSM21} task-based study and are specific to its chart types, tasks, privacy mechanism/parameters, and datasets.}\added{The utility-degradation patterns identified in prior work are specific to the chart types, channel configurations, tasks, DP mechanisms, privacy parameters, datasets, and study methods used in those studies.} \deleted{The candidate hypothesis we offer, while consistent with available evidence , has not been tested in a controlled experiment that isolates encoding properties from confounds.}\added{The channel--read--mechanism hypothesis we propose, while consistent with available evidence, has not been tested in a controlled experiment that isolates channel/read properties from confounds.} \deleted{Our framework addresses basic visual channels; compound encodings introduce interactions we do not address. Extending to multivariate visualizations, where multiple noisy variables interact, is a significant open challenge~\cite{BCD20,ZSM21}.}

Despite these limitations, our synthesis reveals patterns that are less visible when individual studies are considered in isolation. It shifts attention from ``how much noise can we add?'' to ``how should we \deleted{encode}\added{represent} the noisy data, and how does that choice shape disclosure framing, mechanism selection, and interpretation?'' This is a question visualization researchers are uniquely positioned to answer. Panavas et al.'s~\cite{PSD+25} practitioner-interviews 
reveal that practitioners select chart types for different pipeline stages without an established framework; their interviewees express a desire for empirically grounded design guidance. \deleted{We hope this work motivates the systematic, channel-aware research that such guidance requires.}\added{Developing empirically grounded, channel--task-aware design principles is therefore essential for making DP visualizations both private and usable. Such guidance is essential if DP visualization is to move from isolated chart-specific findings toward reusable design principles, and may also inspire new DP mechanisms that better preserve visualization utility.}

\acknowledgments{
This work is supported in part by the National Science Foundation under grants CNS-2144684.  The opinions, findings, and conclusions or recommendations expressed in this material are those of the authors and do not necessarily reflect the views of the sponsors.
}

\bibliographystyle{abbrv-doi}

\bibliography{main}
\end{document}